\documentclass[epj]{svjour}
\usepackage{amsfonts}
\usepackage{graphicx}
\usepackage{amsmath}
\usepackage{amssymb}

\begin{document}

\title{Non-classicalities via perturbing local unitary operations}
\author{Jun Zhang \and Yang Zhang \and Shao-xiong Wu \and Chang-shui Yu%
\thanks{%
quaninformation@sina.com; ycs@dlut.edu.cn} }
\institute{School of Physics and Optoelectronic Technology, Dalian University of
Technology, Dalian 116024, China}
\date{Received: date / Revised version: date}

%


%
%

%
\abstract{
We study the nonclassical correlations in a two-qubit state by the perturbing local unitary operation method. We find that the definitions of various non-classicalities including quantum discord (QD), measurement-induced nonlocality (MIN) and so on usually do not have a unique definition when expressed as the perturbation of local unitary operations, so a given non-classicality can lead to different definitions of its dual
non-classicality. In addition, it is shown that QD and MIN are not the
corresponding dual expressions in a simple set of unitary operations, even
though they are in their original definitions. In addition, we also consider
the non-classicalities in general $2\otimes d$ dimensional systems.
 } 
\maketitle
\section{Introduction}

Quantum correlation is one of the most intriguing features of quantum
mechanics and plays an important role in quantum information. The
quantification of quantum correlation has attracted much attention in recent
times. Among the many measures of quantum correlation such as entanglement
\cite{rmp}, quantum discord \cite{zong,jihe}, the information deficit \cite%
{deficit}, the measurement induced nonlocality \cite{MIN} and so on,
entanglement and quantum discord have been more extensively researched. For
example, entanglement has been investigated extensively and intensively over
the last two decades \cite{rmp} and quantum discord seems to be attracting
increasing interest. However, quantum entanglement and quantum discord are
different, not only in that quantum discord can be present in separable
states, but also in that it can be increased in local operations and
classical communications \cite{pro}. Quantum entanglement has been
identified as an important physical resource in quantum information
processing tasks (QIPTs), but it has been found that some QIPTs without any
entanglement could also display quantum advantages if there exists quantum
discord \cite{DQ,DQC1,D1,D2}. This could shed new light on the role of
quantum discord in quantum computing, hence it could be one of the main
reasons for the recent widespread research on quantum discord in various
fields such as dynamic evolution \cite{power,open,open1,haha,sudden},
Maxwell's demon \cite{demons}, relativistic effects in quantum information
theory \cite{QIT,Q1}, quantum phase transitions \cite%
{phase,phase1,phase2,phase3}, biological systems \cite{thermodynamics} and
so on.

The original definition of quantum discord is information-theoretical \cite%
{Vedral,Olliver}. However, the analytic expression is only available for
some special states \cite{M,S,A,Y,yb,shang,shang1,you}. For this reason, the
geometric version of quantum discord (GD) based on distance measurements was
introduced for a two-qubit system in 2010 \cite{jihe}. It provided a better
method for analytically evaluating $2\otimes d$ dimensional states \cite{2d}%
. Recently, it has been pointed out that the Frobenius norm is unable to
account for experimentally observed contractilities \cite{pro,fanshu},
namely, geometric discord will be increased under a non-unitary evolution
that is described by a completely positive local operation \cite{pro,hu}. In
fact, there are two ways to define distances in vector spaces: 1) by
considering some properly defined norm and the metrics that it induces; 2)
by considering a proper metric, not directly related to a norm, such as the
Fubini-Study metrics and the Bures metrics in classical cases \cite{FS}. In
general, these metrics are, by construction, Riemannian and contractive,
while problems can arise when one considers case 1). Indeed, in
finite-dimensional vector spaces, metrics are induced starting from the
Schatten p-norms, that are the finite-dimensional precursors of the general
p-norms in $L_{p}$ spaces. The case $p=1$ corresponds to the trace norm
(contractive but non-Riemannian); the case $p=2$ corresponds to the
Hilbert-Schmidt norm (not even contractive); the case $p=\infty $
corresponds to the $sup$ norm. Only norms with $p<2$ are contractive. Thus,
all\ kinds of distance measurements of quantum correlations have advantages
and disadvantages \cite{shu}. This way of thinking about quantum correlation
has been systematically studied in Ref. \cite{lp}. Some other definitions of
quantum correlation considering the contractivity have also been proposed
\cite{Xiudis,Lina,prl}. At the same time, some works attempt to compute
distance measures of quantum correlations based on the trace norm \cite%
{trace,dtd}.

In addition, different definitions related to quantum correlations have also
been developed to different extents. For example, quantum discord with
two-side measurements has been considered \cite{47}; the measurement-induced
nonlocality (MIN) is introduced by considering the maximal distance between
the original and the measured density matrices \cite{MIN}. In particular,
geometric quantum discord (even entanglement \cite{76,84}) has been shown to
be redefined via the perturbation of local unitary operations, which also
offered an alternative understanding of quantum discord \cite{LUO,LUOG,fu}.
Thus it is natural to ask whether the MIN can be reconstructed by similar
consideration of the perturbation under local unitary operations and whether
the perturbation of local unitary operations can lead to other interesting
phenomena?

In this paper we consider non-classicalities through perturbation under
local unitary operations. We divide the set of unitary operations into
several sets which are made up of the traceless unitary operations, the
cyclic unitary operations, the special unitary operations and the general
unitary operations, respectively. It is well known that GD and MIN are dual
definitions with respect to local measurements \cite{MIN,LUO,LUOG}. However, we first find that, even though GD can be redefined as the minimal distance
between the state of interest and the one that is perturbed by the local
unitary operations which are in the set of cyclic unitary operations, MIN
cannot be reached by the dual definition (maximization) in the same set. On
the contrary, it will arrive at a new quantity which we call the generalized
MIN (GMIN) in contrast to MIN. Second, the definition of GD based on the
perturbation under local unitary operations are not unique. The minimization
of the distance on both the traceless unitary operation set and the cyclic
unitary operation set can lead to GD. Finally, we also find that there is no
simple unitary operation set (but a relatively complex special set) that is
directly related to the MIN and GD by the dual optimization.

This paper is organized as follows. In Sec. II, we give a brief
classification of the unitary operators used in this work. In Sec. III, we
give the expressions for various non-classicalities based on different sets
of unitary operations. In Sec. IV, we study the connection between GD, MIN
and GMIN. In Sec. V, we expand our results to the $2\otimes d$ dimensional
quantum systems and highlight our conclusions.

\section{Sets of unitary operations}

Let us consider the unitary operations in a 2-dimensional Hilbert space. Any
unitary $U$ can be written up to a constant phase as%
\begin{equation}
U=n_{0}I_{2}+i\mathbf{n}\cdot \mathbf{\sigma },  \label{1}
\end{equation}%
where $I_{n}$ denotes the n-dimensional identity, $n_{0}\in \mathbf{%
\mathbb{R}
,n}=(n_{1},n_{2},n_{3})\in \mathbf{%
\mathbb{R}
}^{3}$ and $\mathbf{\sigma }=(\sigma _{1},\sigma _{2},\sigma _{3})$ with $%
\sigma _{i}$ the Pauli matrices. The unitary property of $U$ requires
\begin{equation}
\sum_{k=0}^{3}n_{k}^{2}=1.
\end{equation}%
Thus, based on the different parameters, one can divide the set of unitary
matrices into different subsets. At first, we would like to use $S_{A}$ to
denote a general unitary matrix given in Eq. (1). If $n_{0}=0$, one will
find that $U$ is traceless. This traceless unitary operator set is denoted
by $S_{T}$. Now let us consider an arbitrary two-qubit state $\rho _{AB}$ in
the Bloch representation as \cite{fano}%
\begin{equation}
\rho _{AB}=\frac{1}{4}[I_{4}+\left( \mathbf{r\cdot \sigma }_{A}\right)
\otimes I_{2}+I_{2}\otimes \left( \mathbf{s\cdot \sigma }_{B}\right)
+\sum\limits_{i,j}T_{ij}\sigma _{i}\otimes \sigma _{j}],
\end{equation}%
where $\mathbf{r},\mathbf{s}$ and $T_{ij}$ are the local Bloch vectors and
the correlation tensor, respectively. Then, the reduced density matrix can
be given by
\begin{equation}
\rho _{A}=\mathrm{Tr}_{B}\rho _{AB}=\frac{1}{2}\left(
\begin{array}{cc}
1+r_{3} & r_{1}-ir_{2} \\
r_{1}+ir_{2} & 1-r_{3}%
\end{array}%
\right) .  \label{3}
\end{equation}%
Suppose a unitary operation $U$ that is performed on $\rho _{A}$ satisfies%
\begin{equation}
\left[ U,\rho _{A}\right] =0,  \label{4}
\end{equation}%
then such unitary operations $U$ will compose a cyclic unitary set $S_{C}$
subject to $\rho _{A}$. In addition, let us use $S_{S}$ to describe the set
made up of some special unitary operations which will be given in Theorem $3$
in the following. Thus, based on these unitary operator sets, we will give
the different formulations of the non-classicalities in the next section.
Before proceeding, we would like to give a lemma that is very useful in this
context.

\textit{Lemma 1.} Eq. (\ref{4}) is equivalent to%
\begin{equation}
c_{1}\mathbf{n}=c_{2}\mathbf{r},\text{ }c_{i}\in \mathbf{%
\mathbb{R}
},
\end{equation}

\textit{Proof.} Inserting Eq. (\ref{1}) and Eq. (\ref{3}) into Eq. (\ref{4}%
), one will arrive at
\begin{gather}
\left( I_{2}+\mathbf{r}\cdot \mathbf{\sigma }\right) \left( n_{0}I_{2}+i%
\mathbf{n}\cdot \mathbf{\sigma }\right) =\left( n_{0}I_{2}+i\mathbf{n}\cdot
\mathbf{\sigma }\right) \left( I_{2}+\mathbf{r}\cdot \mathbf{\sigma }\right)
\notag \\
\Rightarrow \left( \mathbf{r}\cdot \mathbf{\sigma }\right) \left( \mathbf{n}%
\cdot \mathbf{\sigma }\right) -\left( \mathbf{n}\cdot \mathbf{\sigma }%
\right) \left( \mathbf{r}\cdot \mathbf{\sigma }\right) =0  \notag \\
\Rightarrow \mathbf{r}\times \mathbf{n}=0.
\end{gather}%
which shows that there exist real constants $c_{1}$ and $c_{2}$ such that $%
c_{1}\mathbf{n}=c_{2}\mathbf{r}$.

\section{\protect\bigskip Non-classicalities based on the perturbation under
local unitary operations}

\subsection{Non-classicalities via perturbation}

At first, we would like to suppose a local unitary operator $U_{A}$ is
applied to the subsystem $A$ of the state $\rho _{AB}$, giving the final
state:
\begin{equation}
\varrho _{AB}=(U_{A}\otimes I_{2})\rho _{AB}(U_{A}^{\dagger }\otimes I_{2}).
\end{equation}%
Generally, $\varrho _{AB}\neq \rho _{AB}$. So we define%
\begin{equation}
D(\rho _{AB},U_{A}):=\left\Vert \rho _{AB}-\varrho _{AB}\right\Vert ^{2},
\label{7}
\end{equation}%
where $\left\Vert X\right\Vert =\sqrt{\mathrm{Tr}XX^{\dagger }}$ denotes the
Frobenius norm of the matrix $X$. $D(\rho _{AB},U_{A})$ is obviously the
distance between the two states before and after the local unitary
operation. Considering the extremisation of the distance $D(\rho
_{AB},U_{A}) $ by the optimization of different unitary operator sets, one
can define
\begin{equation}
D_{S_{k}}(\rho _{AB}):=\max\limits_{U_{A}\in S_{k}}D(\rho _{AB},U_{A}).
\end{equation}%
which means the maximal distance between the original and the transformed
states induced by the local unitary operators belonging to the corresponding
unitary operator set $S_{k}$ with $k=A,T,C$ or $S$ , and
\begin{equation}
\tilde{D}_{S_{k}}(\rho _{AB}):=\min\limits_{U_{A}\in S_{k}}D(\rho
_{AB},U_{A}).
\end{equation}%
This is the minimal distance that opposes $D_{S_{k}}(\rho _{AB})$. With
these definitions, we will arrive at the following theorem.

\textit{Theorem.}1. For any a two-qubit state $\rho _{AB}$, \ \
\begin{eqnarray}
D_{S_{A}}(\rho _{AB}) &=&D_{S_{T}}(\rho _{AB})=\lambda _{1}+\lambda _{2}, \\
\tilde{D}_{S_{A}}(\rho _{AB}) &=&\tilde{D}_{S_{C}}(\rho _{AB})=0, \\
D_{S_{C}}(\rho _{AB}) &=&D_{MIN}(\rho _{AB}), \\
\tilde{D}_{S_{T}}(\rho _{AB}) &=&D_{GD}(\rho _{AB}), \\
D_{GD}(\rho _{AB}) &=&\mathrm{Tr}A-\lambda _{1}, \\
D_{MIN}(\rho _{AB}) &=&\left\{
\begin{array}{cc}
\mathrm{Tr}TT^{T}-\frac{1}{\left\Vert \mathbf{r}\right\Vert ^{2}}\mathbf{r}%
^{T}TT^{T}\mathbf{r}, & \mathbf{r}\neq 0 \\
\mathrm{Tr}TT^{T}-\lambda _{3}, & \mathbf{r}=0%
\end{array}%
\right. ,
\end{eqnarray}%
where
\begin{equation}
A=\mathbf{rr}^{T}+TT^{T}.  \label{16}
\end{equation}%
and $\lambda _{i}$ is the eigenvalues of $A$ in decreasing order.

\textit{Proof}. $D(\rho _{AB},U_{A})$ can be explicitly given by
\begin{equation}
D(\rho _{AB},U_{A})=\left\Vert \rho _{AB}-\varrho _{AB}\right\Vert ^{2}=2(%
\mathrm{Tr}\rho _{AB}^{2}-\mathrm{Tr}\rho _{AB}\varrho _{AB}).  \label{17}
\end{equation}%
Some simple calculations reveal that%
\begin{eqnarray}  \label{182}
\mathrm{Tr}\rho _{AB}^{2} &=&\frac{1}{4}(2+\left\Vert \mathbf{r}\right\Vert
^{2}+\left\Vert \mathbf{s}\right\Vert ^{2}+\sum\limits_{i,j}T_{ij}^{2}),
\label{181} \\
\mathrm{Tr}\rho _{AB}\varrho _{AB} &=&\frac{1}{4}(2+\left\Vert \mathbf{s}%
\right\Vert ^{2}+\mathrm{Tr}A+2\mathbf{n}A\mathbf{n}^{T}-2\mathrm{Tr}%
A\left\Vert \mathbf{n}\right\Vert ^{2}).  \notag \\
&&  \label{182w}
\end{eqnarray}%
By substituting Eqs. (\ref{181},\ref{182w}) into Eq. (\ref{17}), one will
obtain%
\begin{eqnarray}
&&D(\rho _{AB},U_{A})  \notag \\
&=&2[\frac{1}{4}(\left\Vert \mathbf{r}\right\Vert
^{2}+\sum\limits_{i,j}T_{ij}^{2})-\frac{1}{4}(\mathrm{Tr}A+2\mathbf{n}A%
\mathbf{n}^{T}-2\mathrm{Tr}A\left\Vert \mathbf{n}\right\Vert ^{2}]  \notag \\
&=&\mathbf{n}\left( \mathrm{Tr}AI_{3}-A\right) \mathbf{n}^{T}.  \label{19}
\end{eqnarray}%
with $A\ $ given by Eq. (\ref{16})$.$

$(1)$ If $U_{A}\in S_{A}$, one will find that the calculation of $\tilde{D}%
_{S_{A}}(\rho _{AB})$ is trivial, because one can choose $U_{A}=I_{2}$ such
that $\tilde{D}_{S_{A}}(\rho _{AB})=0$. In addition, Eq. (\ref{19}) can be
rewritten as
\begin{eqnarray}
D(\rho _{AB},U_{A}) &=&\left\Vert \mathbf{n}\right\Vert ^{2}\cdot \frac{%
\mathbf{n}}{\left\Vert \mathbf{n}\right\Vert }\left( \mathrm{Tr}%
AI_{3}-A\right) \frac{\mathbf{n}^{T}}{\left\Vert \mathbf{n}\right\Vert }
\notag \\
&\leq &\lambda _{\max }\left\Vert \mathbf{n}\right\Vert ^{2}  \label{21} \\
&\leq &\lambda _{\max }=\lambda _{1}+\lambda _{2},  \label{20}
\end{eqnarray}%
where $\lambda _{\max }$ denotes the sum of\ the two maximal eigenvalue of $%
A $. In addition, the equality in Eq. (\ref{21}) holds when $\frac{\mathbf{n}}{%
\left\Vert \mathbf{n}\right\Vert }$ is chosen as the eigenvector of $A$
corresponding to its maximal eigenvalue and the equality in Eq. (\ref{20}) is
satisfied if we let $n_{0}=0$.

(2) If $U_{A}\in S_{T}$, we will obtain $n_{0}=0$ and $%
\sum_{k=1}^{3}n_{k}^{2}=1.$ Thus the bounds of $D(\rho ,U_{A})$ can be given
as follows
\begin{eqnarray}
\lambda _{2}+\lambda _{3} &=&\lambda _{\min }  \notag \\
&\leq &\mathbf{n}\left( \mathrm{Tr}AI_{3}-A\right) \mathbf{n}^{T}  \notag \\
&\leq &\lambda _{\max }=\lambda _{1}+\lambda _{2},  \label{22}
\end{eqnarray}%
where $\lambda _{\min }$ and $\lambda _{\max }$ are the sum of\ the two
minimal and the sum of\ the two maximal eigenvalues, respectively, and the
equality in Eq. (\ref{22}) holds if $\mathbf{n}$ is the corresponding
eigenvector. Thus we have $D_{S_{T}}(\rho _{AB})=\lambda _{1}+\lambda _{2}$
and $\tilde{D}_{S_{T}}(\rho _{AB})=\lambda _{2}+\lambda _{3}$.

(3) If $U_{A}\in S_{C}$, we have $c_{1}\mathbf{n}=c_{2}\mathbf{r}$, which
implies three cases:\ a) $\mathbf{n}=0$; b) $\mathbf{r}=0$\textbf{; }c) $%
\mathbf{n}=c\mathbf{r}$, $c\neq 0$. In this case, we can easily find that $%
\tilde{D}_{S_{C}}(\rho _{AB})=0$ if we choose $\mathbf{n}=0$. However, if
we calculate $D_{S_{C}}(\rho _{AB})$, we have to consider whether or not $\mathbf{r}=0$. If b) is satisfied, we will have
\begin{eqnarray}
&&\mathbf{n}\left( \mathrm{Tr}AI_{3}-A\right) \mathbf{n}^{T}  \notag \\
&=&\mathrm{Tr}TT^{T}\left\Vert \mathbf{n}\right\Vert ^{2}-\mathbf{n}TT^{T}%
\mathbf{n}^{T}  \notag \\
&\leq &\mathrm{Tr}TT^{T}-\tilde{\lambda}_{3},  \label{23}
\end{eqnarray}%
with $\tilde{\lambda}_{3}$ the minimal eigenvalue of $TT^{T}$. The equality in
Eq. (\ref{23}) is satisfied when $\mathbf{n}$ is the eigenvector of $TT^{T}$
corresponding to $\tilde{\lambda}_{3}$. If $\mathbf{r}\neq 0$, c) must be
satisfied, so we will arrive at
\begin{eqnarray}
&&\mathbf{n}\left( \mathrm{Tr}AI_{3}-A\right) \mathbf{n}^{T}  \notag \\
&=&\left\Vert \mathbf{n}\right\Vert ^{2}\left( \mathrm{Tr}A-\frac{\mathbf{n}%
}{\left\Vert \mathbf{n}\right\Vert }A\frac{\mathbf{n}^{T}}{\left\Vert
\mathbf{n}\right\Vert }\right)  \notag \\
&=&c^{2}\left\Vert \mathbf{r}\right\Vert ^{2}\left( \mathrm{Tr}A-\frac{%
\mathbf{r}}{\left\Vert \mathbf{r}\right\Vert }A\frac{\mathbf{r}^{T}}{%
\left\Vert \mathbf{r}\right\Vert }\right)  \notag \\
&\leq &\mathrm{Tr}A-\frac{\mathbf{r}}{\left\Vert \mathbf{r}\right\Vert }A%
\frac{\mathbf{r}^{T}}{\left\Vert \mathbf{r}\right\Vert }.  \label{24}
\end{eqnarray}%
The inequality in Eq. (\ref{24}) comes from $\left\Vert \mathbf{n}%
\right\Vert ^{2}\leq 1$. Eq. (\ref{23}) and Eq. (\ref{24}) are just the MIN
introduced in Ref. \cite{MIN}. The proof is completed.

\subsection{Generalized measurement-induced nonlocality}

From the above theorem, we find that the optimization on the traceless
unitary transformations or the all unitary matrices can lead to a new
quantity which can be rewritten as
\begin{equation}
D_{GMIN}(\rho _{AB})=\lambda _{1}+\lambda _{2}.
\end{equation}%
where $\lambda _{i}$ is the eigenvalues of $A=\mathbf{rr}^{T}+TT^{T}$ in
decreasing order. Compared with MIN, we would like to call it the
generalized measurement-induced nonlocality (GMIN). In order to give an
explicit understanding, we would like to introduce several fundamental
properties here.

\textit{Corollary. }1. For a pure two-qubit state,
\begin{equation}
D_{GMIN}(\left\vert \psi \right\rangle _{AB})=2.
\end{equation}

\textit{Proof}. Since, for any bipartite pure state $\left\vert \psi
\right\rangle _{AB}$, $\left\vert \psi \right\rangle _{AB}$ can be given in a
Schmidt decomposition as%
\begin{equation}
\left\vert \psi \right\rangle _{AB}=\sigma _{1}\left\vert 00\right\rangle
+\sigma _{2}\left\vert 11\right\rangle ,
\end{equation}%
where $\sigma _{1}^{2}+\sigma _{2}^{2}=1$, one can obtain the Bloch
vector $\mathbf{s}=(0,0,\sigma _{1}^{2}-\sigma _{2}^{2})^{T}$ and the
correlation tensor
\begin{equation}
T=\left(
\begin{array}{ccc}
2\sigma _{1}\sigma _{2} & 0 & 0 \\
0 & -2\sigma _{1}\sigma _{2} & 0 \\
0 & 0 & 1%
\end{array}%
\right) ,
\end{equation}%
so we have
\begin{equation}
A=\mathbf{ss}^{T}+TT^{T}=\left(
\begin{array}{ccc}
4\sigma _{1}^{2}\sigma _{2}^{2} & 0 & 0 \\
0 & 4\sigma _{1}^{2}\sigma _{2}^{2} & 0 \\
0 & 0 & 1+(\sigma _{1}^{2}-\sigma _{2}^{2})^{2}%
\end{array}%
\right) .
\end{equation}%
Therefore, the sum of the two maximal eigenvalues is $2$.

\textit{Corollary. 2.} For any bipartite product state $\rho =\rho
_{1}\otimes \rho _{2},$ $D_{GMIN}(\rho )=(4\mathrm{Tr}\rho _{1}^{2}-2)(4%
\mathrm{Tr}\rho _{2}^{2}-1)$. If $\rho _{1}$ is an identity, $D_{GMIN}(\rho
)=0$.

\textit{Proof}. The generic single-qubit can be written as%
\begin{eqnarray}
\rho _{1} &=&\frac{1}{2}(1+\mathbf{x}\cdot \mathbf{\sigma }),  \label{123} \\
\rho _{2} &=&\frac{1}{2}(1+\mathbf{y}\cdot \mathbf{\sigma }).  \label{1234}
\end{eqnarray}%
with Bloch vectors $x=(x_{1},x_{2},x_{3})^{T}$ and $%
y=(y_{1},y_{2},y_{3})^{T} $. From here, we will have the bipartite product
state
\begin{equation}
\rho =\frac{1}{4}[I_{4}+\left( \mathbf{x\cdot \sigma }\right) \otimes
I_{2}+I_{2}\otimes \left( \mathbf{y\cdot \sigma }\right)
+\sum\limits_{i,j}x_{i}y_{j}\sigma _{i}\otimes \sigma _{j}],
\end{equation}%
Through Theorem 1, the matrix $A$ will be given by $\mathbf{xx}^{T}+%
\mathbf{xy}^{T}\mathbf{yx}^{T}$. Therefore, the sum of the two maximal
eigenvalues is $\mathbf{x}^{T}\mathbf{x}(1\mathbf{+y}^{T}\mathbf{y}).$ Using
Eqs. (\ref{123},\ref{1234}), one can obtain%
\begin{eqnarray}
\mathrm{Tr}\rho _{1}^{2} &=&\frac{1}{4}(2+\mathbf{x}^{T}\mathbf{x}), \\
\mathrm{Tr}\rho _{2}^{2} &=&\frac{1}{4}(2+\mathbf{y}^{T}\mathbf{y}).
\end{eqnarray}%
It is straightforward $D_{GMIN}(\rho )=(4\mathrm{Tr}\rho _{1}^{2}-2)(4%
\mathrm{Tr}\rho _{2}^{2}-1)$. If $\rho _{1}$ is an identity(i.e.$%
\mathbf{x}=0$), a simple calculation reveals that $D_{GMIN}(\rho )=0.$

\subsection{Relations between various non-classicalities}

From the results given in Theorem 1, we can see that if the optimized
local unitary operations are in the set of traceless unitary
transformations, the minimum of $D(\rho )$ coincides with the GD. In this
sense, it is natural to imagine that the GD is the result of the
perturbation of the traceless unitary transformations. Since the MIN is a
dual definition of quantum discord, it seems that MIN should also be one
result of the perturbation of the traceless unitary transformations.
However, based on our theorem, this is not the case. One can find that the
dual definition of the GD in the framework of perturbation of local unitary
operations is a new quantity, GMIN. From a different perspective, in our
theorem, one can find that the MIN is the result of the perturbation of the
cyclic unitary transformations. Due to the duality of the GMIN and GD, an
intuitive conclusion is that GD should also be the result of the
perturbation of cyclic unitary transformations. However, our theorem tells
us that the minimum induced by the perturbation of the cyclic unitary
transformation is zero instead of the GD. Based on the above analysis, there are two immediate questions. One is what the relation between
GMIN and 0 is, and the other is what the relation between MIN and GD is in
the framework of perturbed unitary matrices. Even though it seems that
GMIN and 0 belong to different kinds of perturbations, they can be unified if
we consider the perturbation of all possible unitary transformations. But
the relation between MIN and GD seems to be rather complex. From Fig. 2,
one can explicitly see that the optimal points corresponding to MIN and GD
are both on the same sphere of traceless unitary transformations, and one might naturally ask whether there exists a simple set such as a circle
on the sphere connecting the two optimal points such that the two points are
still optimal in the sense of the optimization on the circle.
Unfortunately, our following theorem shows us that, generally, such a circle
does not exist. Of course, at any rate, we can always find such a set despite the complexity. These relations are depicted in
Fig. 1.

\begin{figure}[tbp]
\centering
\includegraphics[width=1\columnwidth]{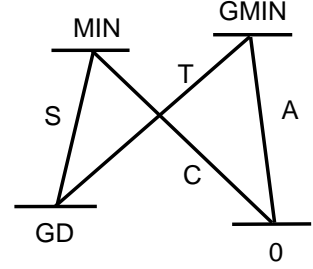}
\caption{The relationship between GD, MIN and GMIN. $A$ stands for the set of all unitary operators, $C$ stands for the cyclic unitary operator set, $T$
stands for the traceless unitary operator set and $S$ stands for the
special unitary operator set. The horizontal lines denote the values of GD,
MIN and GMIN, respectively, and the solid lines connecting them mean the
values they connect can be attained by the perturbation with the corresponding
unitary operations. }
\end{figure}

\textit{Theorem.} 2. On the sphere of traceless unitary transformations for a general two-qubit state $\rho _{AB}$ there is no circle whose dual optimization can reach both MIN and GD.
\begin{figure}[tbp]
\centering
\includegraphics[width=1\columnwidth]{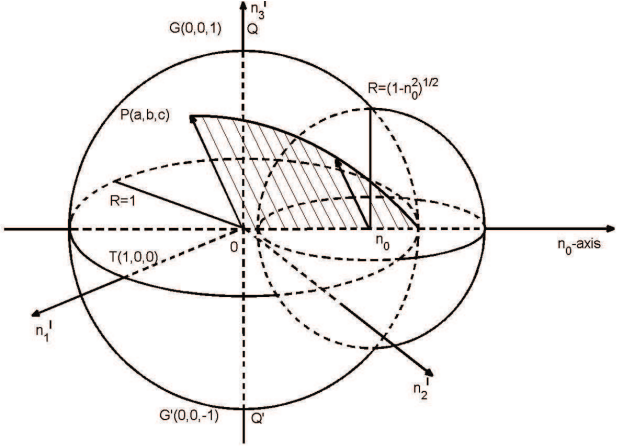}
\caption{The set of unitary operations. The sphere with radius $R=1$ is the
traceless operation set, the small sphere with radius $\protect\sqrt{1-n_0^2}
$ is the set of unitary operations corresponding to a given $n_0$. The solid
line connecting the point $P=(a,b,c)$ and the point $n_0=1$ denotes the
cyclic set for $\left\vert r_{A}\right\rangle \neq 0$. }
\end{figure}

\textit{Proof}. Based on the proof of Theorem 1, one can find that the
optimization problem is given by Eq. (\ref{19}). Now let $A=U\Lambda
U^{\dagger }$ be the eigenvalue decomposition of $A$ and $\sigma _{1}$, $%
\sigma _{2}$ and $\sigma _{3}$ be the eigenvalues in descending order, then
Eq. (\ref{19}) will become
\begin{equation}
D(\rho _{AB},U_{A})=\mathrm{Tr}A-(n_{1}^{\prime 2}\sigma _{1}+n_{2}^{\prime
2}\sigma _{2}+n_{3}^{\prime 2}\sigma _{3}),
\end{equation}%
where $\left\vert n^{\prime }\right\rangle =U\left\vert n\right\rangle ,$ $%
\left\vert r^{\prime }\right\rangle =U\frac{\left\vert r\right\rangle }{%
\left\Vert \mathbf{r}\right\Vert }=(a,b,c)^{T}$ with $a^{2}+b^{2}+c^{2}=1$.
Thus GD is reached at the point $(\pm 1,0,0)$ and MIN is reached at $\left(
a,b,c\right) $. In order to find a pathway to realize the relationship
between MIN and GD, we can use the points $\left( a,b,c\right) $, $(1,0,0)$
(take $(1,0,0)$ for analysis, $(-1,0,0)$ will give similar results) and a
third point $(a^{\prime },b^{\prime },c^{\prime })$ on the sphere to
construct a circle to connect MIN to GD. Here we consider the general case,
so it is implied that the three points are different from each other. Based
on these three points, one can write the equation of the circle as
\begin{equation}
\left\{
\begin{array}{c}
n_{1}^{\prime }+Mn_{2}^{\prime }+Nn_{3}^{\prime }-1=0 \\
n_{1}^{\prime 2}+n_{2}^{\prime 2}+n_{3}^{\prime 2}=1%
\end{array}%
\right. ,
\end{equation}%
where
\begin{equation}
M=\frac{c^{\prime }-c^{\prime }a+ca^{\prime }-c}{c^{\prime }b-cb^{\prime }}%
,N=\frac{b^{\prime }-b^{\prime }a-b+a^{\prime }b}{-c^{\prime }b+cb^{\prime }}%
.
\end{equation}%
Let $\lambda $ and $\mu $ be the Lagrangian multiplier, so the Lagrangian
equation can be written as
\begin{eqnarray}
L &=&\mathrm{Tr}A-(n_{1}^{\prime 2}\sigma _{1}+n_{2}^{\prime 2}\sigma
_{2}+n_{3}^{\prime 2}\sigma _{3})  \notag \\
&&+\lambda (n_{1}^{\prime }+Mn_{2}^{\prime }+Nn_{3}^{\prime }-1)  \notag \\
&&+\mu (n_{1}^{\prime 2}+n_{2}^{\prime 2}+n_{3}^{\prime 2}-1).
\end{eqnarray}%
Taking partial derivative on both sides, one arrives at the following
equations:%
\begin{eqnarray}
\frac{\partial L}{\partial n_{1}^{\prime }} &=&-2n_{1}^{\prime }\sigma
_{1}+2\mu n_{1}^{\prime }+\lambda ,  \label{first1} \\
\frac{\partial L}{\partial n_{2}^{\prime }} &=&-2n_{2}^{\prime }\sigma
_{2}+2\mu n_{2}^{\prime }+\lambda M,  \label{first2} \\
\frac{\partial L}{\partial n_{3}^{\prime }} &=&-2n_{3}^{\prime }\sigma
_{3}+2\mu n_{3}^{\prime }+\lambda N,  \label{first3} \\
\frac{\partial L}{\partial \lambda } &=&n_{1}^{\prime }+Mn_{2}^{\prime
}+Nn_{3}^{\prime }-1,  \label{first4} \\
\frac{\partial L}{\partial \mu } &=&n_{1}^{\prime 2}+n_{2}^{\prime
2}+n_{3}^{\prime 2}-1.  \label{first}
\end{eqnarray}%
Since both $\left( a,b,c\right) $ and $(1,0,0)$ are the extreme points, they
should satisfy the above five questions. Inserting $(a,b,c)$ into these
equations, we arrive at%
\begin{eqnarray}
\mu &=&\frac{-a\sigma _{1}+a^{2}\sigma _{1}+b^{2}\sigma _{2}+c^{2}\sigma _{3}%
}{1-a}, \\
\lambda &=&\frac{2(ab^{2}\sigma _{1}+ac^{2}\sigma _{1}-ab^{2}\sigma
_{2}-ac^{2}\sigma _{3})}{1-a}, \\
N &=&\frac{c[-(-1+a)a\sigma _{1}-b^{2}\sigma _{2}+(-a+a^{2}+b^{2})\sigma
_{3}]}{a[(b^{2}+c^{2})\sigma _{1}-b^{2}\sigma _{2}-c^{2}\sigma _{3}]},
\label{m} \\
M &=&\frac{b[-(-1+a)a\sigma _{1}-c^{2}\sigma _{3}+(-a+a^{2}+b^{2})\sigma
_{2}]}{a[(b^{2}+c^{2})\sigma _{1}-b^{2}\sigma _{2}-c^{2}\sigma _{3}]}.
\label{n}
\end{eqnarray}%
By considering Eq. (\ref{m}), Eq. (\ref{n}) and $a^{\prime 2}+b^{\prime
2}+c^{\prime 2}-1=0,$ one finds that $(a^{\prime },b^{\prime
},c^{\prime })$ is either $(a,b,c)$ or $(1,0,0)$ in the general case, which
contradicts our previous requirement that $(a^{\prime },b^{\prime
},c^{\prime })$ should be different from $(a,b,c)$ and $(1,0,0)$. The proof
is completed.

Since we have shown that there is no such set of unitary operations that
forms a circle on the sphere of traceless unitary operations, the obvious question is: what set of unitary operations does correspond to the dual optimization of MIN and GD. This is given by our next theorem.

\textit{Theorem. 3.} The set $S_{S}$ of the unitary operations corresponding
to the dual optimization of MIN and GD is given by%
\begin{equation}
S_{S}=\left\{ U|\mathrm{Tr}\left\vert \left[ \rho ,\tilde{U}_{C}\right]
\right\vert ^{2}\geq \mathrm{Tr}\left\vert \left[ \rho ,U\right] \right\vert
^{2}\right\} ,
\end{equation}%
where $\tilde{U}_{C}$ is the optimal unitary matrix leading to MIN.

\textit{Proof.} Let $D$ denote the distance between $\rho $ and $\rho
_{f}=(U_{A}\otimes I_{2})\rho _{AB}(U_{A}^{\dagger }\otimes I_{2})$, so
\begin{eqnarray}
D_{MIN}-D &=&(2\mathrm{Tr}\rho ^{2}-2\mathrm{Tr}\rho \tilde{\rho}_{f})-(2%
\mathrm{Tr}\rho ^{2}-2\mathrm{Tr}\rho \rho _{f})  \notag \\
&=&2\mathrm{Tr}\rho \rho _{f}-2\mathrm{Tr}\rho \tilde{\rho}_{f}  \notag \\
&=&2\mathrm{Tr}\rho U\rho U^{\dagger }-2\mathrm{Tr}\rho \tilde{U}_{C}\rho
\tilde{U}_{C}^{\dagger }.  \label{42}
\end{eqnarray}%
Using%
\begin{equation}
\mathrm{Tr}\left\vert \left[ \rho ,U\right] \right\vert ^{2}=2\mathrm{Tr}%
\rho ^{2}-2\mathrm{Tr}\rho U\rho U^{\dagger },
\end{equation}%
Eq. (\ref{42}) can be rewritten as%
\begin{equation}
D_{MIN}-D=\mathrm{Tr}\left\vert \left[ \rho ,\tilde{U}_{C}\right]
\right\vert ^{2}-\mathrm{Tr}\left\vert \left[ \rho ,U\right] \right\vert
^{2}.
\end{equation}%
Since $D_{MIN}$ is the maximum in the set,%
\begin{equation}
\mathrm{Tr}\left\vert \left[ \rho ,\tilde{U}_{C}\right] \right\vert ^{2}\geq
\mathrm{Tr}\left\vert \left[ \rho ,U\right] \right\vert ^{2}.
\end{equation}%
This completes the proof.

\begin{figure}[tbp]
\centering
\includegraphics[width=0.75\columnwidth]{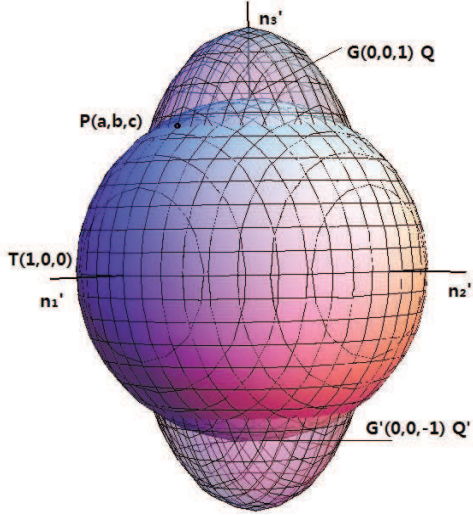}
\caption{The special set of unitary operations. The points between the two
curves on the sphere where the spheroid and the sphere intersect correspond
to the special set of unitary operations that connect MIN and GD. The points
$P,T,Q,Q^{\prime},G,G^{\prime}$ are analyzed in the text.}
\end{figure}

In order to provide an intuitive illustration, we would like to sketch the
different sets of unitary transformations in Fig. 2. Since any unitary
operators on a qubit can be written as Eq. (\ref{1}), one will find that,
given an $n_{0}$, $\mathbf{n}$ characterizes a sphere with radius $\sqrt{%
1-n_{0}^{2}}$. If we let the horizontal axis denote $n_{0}$ in Fig. 2,
with $n_{0}$ changing from $0$ to $1$, the sphere with radius $1$
corresponding to the traceless unitary operators set $S_{T}$ will reduce to
a point at $n_{0}=1$ corresponding to a unit operator. Denoting the point $%
(a,b,c)$ by $P$ on the sphere which corresponds to the optimal unitary
matrix that attains MIN with $\left\vert r_{A}\right\rangle \neq 0.$ Hence with
the change of $n_{0}$, $P$ will undergo a trajectory which is also plotted
in the Fig. 2 and denotes the cyclic unitary operator set $S_{C}$ with $%
\left\vert r_{A}\right\rangle \neq 0.$ If $\left\vert r_{A}\right\rangle =0$%
, $S_{C}$ consistent with $S_{A}$ corresponds to all the spheres with
different $n_{0}$ included. Based on the above calculation, $(0,0,\pm 1)$,
denoted by $Q(Q^{\prime })$, is the optimal point corresponding to MIN with $%
\left\vert r_{A}\right\rangle =0$ which is consistent with the optimal point
$G(G^{\prime })$ corresponding to GMIN. In this case, the set $S_{S}=S_{T}$.
Substituting $(a,b,c)$ and an arbitrary point $(a^{\prime },b^{\prime
},c^{\prime })$ into Eq. (\ref{42}), one will arrive at%
\begin{equation}
\frac{a^{\prime 2}}{\frac{\Delta }{\sigma _{1}}}+\frac{b^{\prime 2}}{\frac{%
\Delta }{\sigma _{2}}}+\frac{c^{\prime 2}}{\frac{\Delta }{\sigma _{3}}}\geq
1,  \label{46}
\end{equation}%
where $\Delta =a^{2}\sigma _{1}+b^{2}\sigma _{2}+c^{2}\sigma _{3}$. Eq. (\ref%
{46}) describes a spheroid and its outer part which is also shown in Fig. 3.
The two curves where the spheroid and the sphere intersect show the
potential optimal points for MIN (generally for $\left\vert
r_{A}\right\rangle \neq 0$). The points between the two curves on the sphere
comprise the set $S_{S}$ with $\left\vert r_{A}\right\rangle \neq 0$. It is
obvious that if the point $P(\left\vert r_{A}\right\rangle )$ serves as the
point of intersection of the curves and the $n_{2}^{\prime }-O-n_{3}^{\prime
}$ plane, there can exist a great circle that connects MIN and GD. If the
point $P$ ($\left\vert r_{A}\right\rangle $) serves as the point of
intersection of the curves and the $n_{1}^{\prime }-O-n_{3}^{\prime }$
plane, there can exist a small circle that connects MIN and GD. Thus it is
also apparent that, in the general case, there is no circle on the sphere that directly relates MIN to GD.

\section{The non-classicalities of $2\otimes d$ dimensional quantum systems.}

In this section we we will discuss the non-classicalites of the qubit-qudit
quantum state. Any $2\otimes d$ dimensional quantum system can be written in
the following form \cite{fano}
\begin{eqnarray}
\rho _{AB} &=&\frac{1}{2d}[\mathbf{I+(r}\cdot \mathbf{\sigma }_{A})\otimes
\mathbf{I+}\sqrt{\frac{d(d-1)}{2}}\mathbf{I}\otimes (\mathbf{s}\cdot \mathbf{%
\sigma }_{B})  \notag \\
&&+\sqrt{\frac{d(d-1)}{2}}\sum\limits_{i,j}T_{ij}\sigma _{i}\otimes \sigma
_{j}].
\end{eqnarray}%
where $\sigma _{A}$ and $\sigma _{B}$ are the generators of $SU(2)$ and $%
SU(d)$, $\mathbf{r},\mathbf{s}$ and $T_{ij}$ are components of the local
Bloch vectors and the correlation tensor, respectively. The final state
after the local unitary perturbation is%
\begin{eqnarray}
\varrho _{AB} &=&\frac{1}{2d}[\mathbf{I+(r}\cdot U_{A}\mathbf{\sigma }%
_{A}U_{A}^{\dagger })\otimes \mathbf{I+}\sqrt{\frac{d(d-1)}{2}}\mathbf{I}%
\otimes (\mathbf{s}\cdot \mathbf{\sigma }_{B})  \notag \\
&&+\sqrt{\frac{d(d-1)}{2}}\sum\limits_{i,j}T_{ij}U_{A}\sigma
_{i}U_{A}^{\dagger }\otimes \sigma _{j}].
\end{eqnarray}%
Thus we arrive at the following results:

\textit{Theorem. 4.} For any $2\otimes d$ dimensional quantum systems $\rho
_{AB}$, \ \
\begin{eqnarray}
D_{S_{A}}(\rho _{AB}) &=&D_{S_{T}}(\rho _{AB})=\frac{4}{d^{2}}(\lambda
_{1}+\lambda _{2}), \\
\tilde{D}_{S_{A}}(\rho _{AB}) &=&\tilde{D}_{S_{C}}(\rho _{AB})=0, \\
D_{S_{C}}(\rho _{AB}) &=&\frac{2(d-1)}{d}D_{MIN}(\rho _{AB}), \\
\tilde{D}_{S_{T}}(\rho _{AB}) &=&\frac{4}{d^{2}}D_{GD}(\rho _{AB}), \\
D_{GD}(\rho _{AB}) &=&\mathrm{Tr}A-\lambda _{1}, \\
D_{MIN}(\rho _{AB}) &=&\left\{
\begin{array}{cc}
\mathrm{Tr}TT^{T}-\frac{1}{\left\Vert \mathbf{r}\right\Vert ^{2}}\mathbf{r}%
^{T}TT^{T}\mathbf{r}, & \mathbf{r}\neq 0 \\
\mathrm{Tr}TT^{T}-\lambda _{3}, & \mathbf{r}=0%
\end{array}%
\right. .
\end{eqnarray}%
where $A=\mathbf{rr}^{T}+\frac{d(d-1)}{2}TT^{T}$\ and $\lambda _{i}$ is the
eigenvalues of $A$ in decreasing order.

\textit{Proof}. $D(\rho _{AB},U_{A})$ can be explicitly given by
\begin{equation}
D(\rho _{AB},U_{A})=\left\Vert \rho _{AB}-\varrho _{AB}\right\Vert ^{2}=2(%
\mathrm{Tr}\rho _{AB}^{2}-\mathrm{Tr}\rho _{AB}\varrho _{AB}).  \label{shang}
\end{equation}%
After the simple calculations, one has%
\begin{eqnarray}
\mathrm{Tr}\rho _{AB}^{2} &=&\frac{1}{4d^{2}}\mathrm{Tr}[\mathbf{I+\mathbf{rr%
}}^{T}\mathbf{\sigma }_{A}\mathbf{\sigma }_{A}\otimes \mathbf{I}  \notag \\
&&+\frac{d(d-1)}{2}\sum\limits_{i,j}\sum\limits_{m,n}T_{ij}T_{mn}\sigma
_{i}\sigma _{m}\otimes \sigma _{j}\sigma _{n}]  \notag \\
&&\mathbf{+}\frac{d(d-1)}{2}\mathbf{I}\otimes \mathbf{\mathbf{ss}}^{T}%
\mathbf{\sigma }_{B}\mathbf{\sigma }_{B},  \label{t1} \\
\mathrm{Tr}\rho _{AB}\varrho _{AB} &=&\frac{1}{4d^{2}}\mathrm{Tr}[\mathbf{I+%
\mathbf{rr}}^{T}\mathbf{\sigma }_{A}U_{A}\mathbf{\sigma }_{A}U_{A}^{\dagger
}\otimes \mathbf{I}  \notag \\
&&\mathbf{+}\frac{d(d-1)}{2}\sum\limits_{i,j}\sum\limits_{m,n}T_{ij}T_{mn}%
\sigma _{i}U_{A}\sigma _{m}U_{A}^{\dagger }\otimes \sigma _{j}\sigma _{n}]
\notag \\
&&+\frac{d(d-1)}{2}\mathbf{I}\otimes \mathbf{\mathbf{ss}}^{T}\mathbf{\sigma }%
_{B}\mathbf{\sigma }_{B}.  \label{t2}
\end{eqnarray}%
Substituting Eqs. (\ref{t1},\ref{t2}) into Eq. (\ref{shang}), one obtains%
\begin{equation}
D(\rho _{AB},U_{A})=\frac{4}{d^{2}}\mathbf{n}(\mathrm{Tr}AI-A)\mathbf{n}^{T}
\label{61}
\end{equation}%
where $A\ $is given by $A=\mathbf{rr}^{T}+\frac{d(d-1)}{2}TT^{T}.$

(1) It is trivial to show that $\tilde{D}%
_{S_{A}}(\rho _{AB})=0$ when $U_{A}\in S_{A}$. In addition, Eq. (\ref{61}) can be
rewritten as
\begin{eqnarray}
D(\rho _{AB},U_{A}) &=&\frac{4}{d^{2}}\left\Vert \mathbf{n}\right\Vert
^{2}\cdot \frac{\mathbf{n}}{\left\Vert \mathbf{n}\right\Vert }\left( \mathrm{%
Tr}AI-A\right) \frac{\mathbf{n}^{T}}{\left\Vert \mathbf{n}\right\Vert }
\notag \\
&\leq &\frac{4}{d^{2}}\lambda _{\max }\left\Vert \mathbf{n}\right\Vert ^{2}
\label{621} \\
&\leq &\frac{4}{d^{2}}\lambda _{\max }=\frac{4}{d^{2}}(\lambda _{1}+\lambda
_{2}),  \label{62}
\end{eqnarray}%
where $\lambda _{\max }$ denotes the sum of the two maximal eigenvalues of $A$%
. In addition, the equality in Eq. (\ref{621}) holds if and only if $\frac{%
\mathbf{n}}{\left\Vert \mathbf{n}\right\Vert }$ is chosen as the eigenvector
of $A$ corresponding to its maximal eigenvalue. And if we let $n_{0}=0$, the
equality in Eq. (\ref{62}) will be reached.

(2) When $U_{A}\in S_{T}$, it means that $n_{0}=0$ and $%
\sum_{k=1}^{3}n_{k}^{2}=1.$ Thus the upper and lower bounds on $D(\rho
_{AB},U_{A})$ will be given as follows:
\begin{eqnarray}
\lambda _{2}+\lambda _{3} &=&\lambda _{\min }  \notag \\
&\leq &\frac{4}{d^{2}}\mathbf{n}\left( \mathrm{Tr}AI-A\right) \mathbf{n}^{T}
\notag \\
&\leq &\frac{4}{d^{2}}\lambda _{\max }=\frac{4}{d^{2}}(\lambda _{1}+\lambda
_{2}),  \label{63}
\end{eqnarray}%
where $\lambda _{\min }$ and $\lambda _{\max }$ are the sum of\ the two
minimal and the sum of\ the two maximal eigenvalues of $A$, respectively. If
$\mathbf{n}$ takes the corresponding eigenvector, the equality in Eq. (\ref{63}) holds. In this case, $D_{S_{T}}(\rho _{AB})=\frac{4}{d^{2}}(\lambda
_{1}+\lambda _{2})$ and $\tilde{D}_{S_{T}}(\rho _{AB})=\frac{4}{d^{2}}%
(\lambda _{2}+\lambda _{3})$. This result coincides with the quantum
discord for a qubit-qudit system\cite{2d}.

(3) When $U_{A}\in S_{C}$, we have from Lemma 1 that $c_{1}\mathbf{n}=c_{2}%
\mathbf{r}$, which also implies three cases:\ a) $\mathbf{n}=0$; b) $\mathbf{%
r}=0$\textbf{; }c) $\mathbf{n=}c\mathbf{r}$, $c\neq 0$, just like in the
two-qubit quantum states. In this case, if we choose $\mathbf{n}=0$ we can
easily find that $\tilde{D}_{S_{C}}(\rho _{AB})=0$. However, if we want to
calculate $D_{S_{C}}(\rho _{AB})$, we have to consider whether $%
\mathbf{r}=0$ or not. If b) is satisfied, we will have
\begin{eqnarray}
&&\frac{4}{d^{2}}\mathbf{n}\left( \mathrm{Tr}AI-A\right) \mathbf{n}^{T}
\notag \\
&=&\frac{2(d-1)}{d}(\mathrm{Tr}TT^{T}\left\Vert \mathbf{n}\right\Vert ^{2}-%
\mathbf{n}TT^{T}\mathbf{n}^{T})  \notag \\
&\leq &\frac{2(d-1)}{d}(\mathrm{Tr}TT^{T}-\tilde{\lambda}_{3}),  \label{64}
\end{eqnarray}%
with $\tilde{\lambda}_{3}$ the minimal eigenvalue of $TT^{T}.$ The equality in
Eq. (\ref{64}) is satisfied when $\mathbf{n}$ is the eigenvector of $TT^{T}$
corresponding to $\tilde{\lambda}_{3}$. If $\mathbf{r}\neq 0$, c) has to be
satisfied. Hence, we arrive at
\begin{eqnarray}
&&\frac{4}{d^{2}}\mathbf{n}\left( \mathrm{Tr}AI-A\right) \mathbf{n}^{T}
\notag \\
&=&\frac{4}{d^{2}}\left\Vert \mathbf{n}\right\Vert ^{2}\left( \mathrm{Tr}A-%
\frac{\mathbf{n}}{\left\Vert \mathbf{n}\right\Vert }A\frac{\mathbf{n}^{T}}{%
\left\Vert \mathbf{n}\right\Vert }\right)  \notag \\
&=&\frac{4}{d^{2}}c^{2}\left\Vert \mathbf{r}\right\Vert ^{2}\left( \mathrm{Tr%
}A-\frac{\mathbf{r}}{\left\Vert \mathbf{r}\right\Vert }A\frac{\mathbf{r}^{T}%
}{\left\Vert \mathbf{r}\right\Vert }\right)  \notag \\
&\leq &\frac{2(d-1)}{d}\left( \mathrm{Tr}TT^{T}-\frac{\mathbf{r}}{\left\Vert
\mathbf{r}\right\Vert }TT^{T}\frac{\mathbf{r}^{T}}{\left\Vert \mathbf{r}%
\right\Vert }\right) .  \label{65}
\end{eqnarray}%
The inequality in Eq. (\ref{65}) comes from $\left\Vert \mathbf{n}%
\right\Vert ^{2}\leq 1$. Eq. (\ref{64}) and Eq. (\ref{65}) are just the MIN
introduced in Ref. \cite{MIN}.

\section{Discussions and Conclusion}

Before concluding, we would like to make a brief comparison between the results
of Ref. \cite{LUO} or Ref. \cite{LUOG} and our results. Ref. \cite%
{LUO} and Ref. \cite{LUOG} showed that the geometric measure of quantum
correlations can be defined not only by taking the distance from the state
being considered and its image under a generic local unitary operation, but
also by minimizing this distance over all traceless local unitary
operations. However, in our work we employ the same method, i.e., the
perturbation of local unitary operation, not only to obtain geometric
discord as those in Ref. \cite{LUO} and Ref.\cite{LUOG}, but also to obtain
the dual definitions of discord, i.e., the MIN and GMIN. In particular, we
mainly emphasize that for a given quantum correlation, its dual definition
of quantum correlation is not unique based on the local unitary
perturbations. This has been explicitly illustrated in Fig. 1. Regarding local unitary operations, we have considered various types
of unitary operation sets such as $S_A$, $S_C$, $S_T$ and $S_S$, while
others only considered a few of them.

To sum up, we have studied non-classicalities based on the perturbation
of local unitary operations. We find that both GD and MIN can be
understood in this way. However, even though GD and MIN are the two dual
definitions in the framework of their original definitions, they cannot be
connected by a simple set of unitary operations in the sense of the
perturbation of local unitary operations. On the contrary, it is shown that
they are connected in a strange way by the set $S_{S}$. In addition, we find
that all the non-classicalities described in the paper have no unique
explanation based on the perturbation of the local unitary operations,
which naturally leads to their dual definitions being quite different,
that is, the dual definition is strongly dependent on the perturbing unitary
operation set. We hope this will shed new light on quantum correlations.
Lastly, we would like to say that the traceless local unitary operation is
only sufficient for qubit systems and one must consider the completely
non-degenerate traceless unitary operations for high dimensional systems. In
addition, all the quantities given in this paper are based on the Frobenius
Norm, which is not-contractive. Considering the recent exciting results based on
skew information \cite{prl}, we look forward to attempting to relate local unitary perturbation to skew-information based measures.

\section{Acknowledgement}

This work was supported by the National Natural Science Foundation of China,
under Grant No. 11175033 and by the Fundamental Research Funds of the
Central Universities, under Grant No. DUT12LK42.

%
\bibliographystyle{Style/icpig}
\bibliography{MicroPhys}

\end{document}